\documentclass[universe,article,accept,moreauthors,pdftex]{Definitions/mdpi} 

\firstpage{1} 
\pubvolume{1}
\issuenum{1}
\articlenumber{0}
\pubyear{2024}
\copyrightyear{2024}
\externaleditor{Academic Editor: Firstname\linebreak  Lastname}
\datereceived{31 January 2024} 
\daterevised{7 March 2024} 
\dateaccepted{22 March 2024} 
\datepublished{ } 
\hreflink{https://doi.org/} 

\usepackage{graphicx,epsfig}
\usepackage{lipsum}
\usepackage{amsfonts}
\input{epsf}

\definecolor{falured}{rgb}{0.5, 0.09, 0.09}

\def\be{\begin{equation}}
\def\ee{\end{equation}}
\def\ba{\begin{eqnarray}}
\def\ea{\end{eqnarray}}

\def\msun{M_\odot}

\def\ltsima{$\; \buildrel < \over \sim \;$}
\def\simlt{\lower.5ex\hbox{\ltsima}}
\def\gtsima{$\; \buildrel > \over \sim \;$}
\def\simgt{\lower.5ex\hbox{\gtsima}}
\def\ltgt{$\; \buildrel < \over \geq \;$}
\def\gtlt{\lower.5ex\hbox{\ltgt}}

\def\zsun{Z_\odot}

\Title{Tracking Dusty Cloud Crushed by a Hot Flow}

\TitleCitation{Tracking Dusty Cloud Crushed by a Hot Flow}

\Author{{Svyatoslav Dedikov}$^{\dagger}$\orcidA{} and {Evgenii Vasiliev}$^{*,\dagger}$\orcidB{}}

\AuthorNames{Svyatoslav Dedikov, Evgenii Vasiliev}

\AuthorCitation{{Dedikov, S.;} 
 Vasiliev, E.}

\address[1]{
Lebedev Physical Institute of Russian Academy of Sciences, 53 Leninskiy Ave., Moscow 119991, Russia;
{s.dedikov@asc.rssi.ru} 
}

\corres{\hangafter=1 \hangindent=1.05em \hspace{-0.82em}Correspondence: eugstar@mail.ru}

\firstnote{\hangafter=1 \hangindent=1.05em \hspace{-0.82em}These authors contributed equally to this work.}
\abstract{
The destruction of clouds by strong shocks and hot winds is the key process responsible for the transporting of metals and dust from the ISM to the ICM/IGM, and establishing the multiphase structure in and around galaxies. In this work, we perform a detailed analysis of this process using two different approaches for tracking the cloud material (gas and dust): the so-called ‘colored’ fluid, and the Lagrangian (trace) particles. We find that for the clouds in the hot phase ($T>10^5$ K), the two methods produce significantly different mass fractions and velocities of the cloud material. In contrast, the two methods produce similar results for the clouds that are in the warm/cold phases ($T < 10^5$ K). We find that the Kelvin--Helmholtz instability is suppressed in the warm clouds of size $\sim$100 pc and metallicity $Z\simgt 0.1\zsun$ due to effective gas cooling. This causes a delay in the destruction of such clouds that are interacting with the hot ICM flow. We demonstrate that the dust particles that are evacuated from their ‘parent’ clouds to the hot medium show different dynamics when compared to that of the Lagrangian (trace) particles. Our results indicate that the dust grains swept out to the hot gas are destroyed. 
}


\begin{document}
\

\section{Introduction}

It has been recognized for a long time that interstellar dust can be elevated outside galactic disks into their halos and circumgalactic gas (CGM), and farther away into extragalactic space either by radiation pressure or by supernova-driven winds ({e.g.,}  \citep{Ferrara1990,Shustov1995,Bianchi2005,Sharma2011,Gupta2016}). An interstellar magnetic field seems to inhibit radiation-driven dust outflows as mentioned in \citep{Shchekinov2011}; however, under certain conditions, large-scale magneto-gravitational (Parker) instability can enhance them \citep{Dettmar2001}. Observational evidence of dust's presence in the galactic  halo at heights $h\sim 2$--$5$ kpc were reported first in the galaxy NGC 891 \citep{Howk1997}. Later on, clear traces of dust at 10--12 kpc above the disk plane were confirmed in \citep{Howk1999,Bocchio2016-ngc891}.  Dusty halos extending into the circumgalactic environment at much larger scales, $r\sim$ 60--200 kpc, have been found in two nearby galaxies, NGC 2835 and NGC 3521 \citep{Zaritsky1994}. This result was confirmed in later studies~\citep{Sembach1996,Hodges-Kluck2014,Ruoyi2020}.

More recently, the presence of dust on megaparsec scales, $r\sim 10$~Mpc, with a considerable dust mass, $M\sim 5\times 10^7\msun$, and a dust-to-gas mass fraction of the order of the local ISM, extending from parent galaxies to the intergalactic medium, has been demonstrated in~\citep{Menard2010,Peek2015}. It is clear that the transport of unshielded dust particles over such long distances in a hot environment in a CGM is accompanied by the destruction of the grains via thermal or kinetic sputtering \citep{Draine1979b}. Therefore, it is usually assumed that such transport is carried out via the expelling of dense and cold gas clouds and clumps containing dust {(for} a more recent discussion, see review \citep{Shchekinov2022Univ}).

Estimates of the dust mass fraction in galaxy clusters (GCs) are less certain. Measurements of extinction and reddening reveal the presence of dust up to radii $\sim$$10R_{200}\sim 1$~Mpc with the total dust mass in the intracluster medium (ICM) of $\sim$$10^9~\msun$, i.e, the dust-to-gas mass ratio is around 0.4\% of the local ISM, assuming the ICM mass $\sim$$4\times 10^{13}~\msun$ \citep{Chelouche2007,McGee2010}. On the other hand, estimates of the dust-to-gas mass ratio inferred from the IR observation of GCs only allow an upper limit to be put, which vary for different GC samples from \mbox{$<$$0.015$\% \citep{Gutirrez2017}} to  \mbox{$<$$0.1$\%  \citep{Roncarelli2010,Gutirrez2014}} of the local ISM. However, in both cases, the dust-to-gas mass ratio is considerably lower than in the local ISM, indicating a much stronger degree of dust destruction than is found 
 in the galactic and circumgalactic haloes. The reasons for such differences are not very obvious because the distances between galaxies in a GC of the order of $\sim$$10N_g^{-1/3}$ Mpc are smaller than the sizes of the CGM regions in \citep{Menard2010,Peek2015}; here, $N_g$ is the number of galaxies in the cluster. Numerical analyses of these reasons are important for understanding the life cycle of dust on intergalactic scales.

It is known that naked (unshielded) dust particles are destroyed in a rather dense ($n>10^{-3}$ cm$^{-3}$) and hot ($T>10^6$ K) environment in a CGM and in an ICM gas while traveling on timescales close to the crossing times typical for a CGM and ICM \citep{Shchekinov2022Univ,Polikarpova2017}. Usually, dust expelled out of galaxies is supposed to be confined into denser gas clouds. Under such conditions, dust survives for lengths of time comparable to the clouds' lifetimes that suffered destruction by hydrodynamical instabilities while moving outwards. Once clouds are destructed, dust particles of 0.1 $\upmu$m can survive no longer than $\sim$$100$ Myr. This is, therefore, our primary motivation to consider hydrodynamical destruction of dusty clouds moving through a hot intracluster gas.

The problem we are pursuing here is equivalent to a gas cloud interacting with a wind, with an additional process connected with dust destruction when dust particles come into direct contact with a hot and dense ambient gas. This requires an adequate numerical procedure which would allow tracking an immediate environmental conditions in a gas volume carrying a dust particle.   

To construct such models, one needs to track the cloud material as accurately as possible in order to diminish numerical diffusion of a dust particle into a hot environment. Complications come from different dynamics of gas and dust along with an unavoidable influence of numerical mixing of cold cloud and hot ambient gas while the cloud moves.

Our goal here is to minimize these effects. In several studies, the dynamics of dust has been described by using Lagrangian (trace) particles \citep{Farber2022}. This approach cannot be considered as adequate in general due to dust being inertial and suffering drag force by a gas \citep{Hopkins2016}; its mass can become smaller due to destruction during the evolution. In this paper, we can distinguish the dynamics of Lagrangian and dust particles, and compare them. In Section~\ref{sec-mod}, we describe our model. In Section~\ref{sec-res0}, we consider the tracking methods. In Section~\ref{sec-res}, we describe the dynamics of the 'cloud-wind' 
 interaction. In Section~\ref{sec-con}, we summarize our results.

\section{Model Description} 
\label{sec-mod}

In order to understand the processes of cloud destruction, mass entrainment, and metal enrichment, one should correctly track the material (gas and dust) belonging to a crushed cloud. To do this, such gas or dust is marked by `color', and then the dispersal or mixing of such colored gas or dust is followed \citep{Xu1995,Avillez2002,Farber2022}. Here, we consider two approaches for tracking cloud material: by `colored' fluid and by Lagrangian (trace) particles. We follow the dynamics of dust as inertial particles interacting with ambient gas by drag force. Thus, we can compare the dynamics of cloud material including dust using different tracking~methods.

We consider the interaction between a dusty cloud and a hot gas flowing around it.
We assume that a cloud of atomic hydrogen is put into a hot flow associated with either wind or a shockwave in the ICM. At the initial moment, dust grains are located inside the cloud, whereas no dust in the hot flow is assumed. We are interested in the dynamics of dust particles swept out of the cloud by the hot flow, and their entrainment and destruction during the cloud crushing.

\subsection{Gas and Dust}

Initially, a cloud is in pressure equilibrium with the medium {at rest}.  {During the next step, it becomes involved in a flow behind a shockwave or wind.} The typical time for cloud destruction in an adiabatic case is $t_{cc} = \chi^{1/2} r_{cl,0}/v_{s,w}$ \citep{Klein1994}, where $\chi = \rho_c / \rho_h$ is the cloud overdensity ($\rho_c$ is the density of {the} cloud, {the cloud is assumed uniform}, $\rho_h$ is the density of the hot medium), $r_{cl,0}$ is the cloud radius, and $v_{s,w}$ is the shock/wind velocity. We normalize timescales to the cloud-crushing time and follow our simulations till $t=10t_{cc}$. 

We consider a cloud with $n_c = 0.1$ cm$^{-3}$, $T_c = 10^4$ K embedded into the medium with $n_m = 10^{-3}$ cm$^{-3}$, $T_m = 10^6$ K, i.e., the overdensity is $\chi=100$. The radius of the cloud is equal to 100 pc. The metallicity adopted for both phases equals 0.1 of the solar value. The inflow velocity of a gas $v_s$ is set to 300 km/s. The crushing time is $\sim$3~Myr for these parameters. 
The postshock flow has parameters according to the Rankine--Hugoniot~conditions. 

We assume that dust particles are contained in {the} cloud only. At the initial moment, they have a size of 0.1 $\upmu$m, and the dust-to-gas {mass} ratio is ${\cal D} = 10^{-2}$ for solar metallicity. Dust grains can be destructed by both thermal (in hot gas) and kinetic (due to relative high-velocity motions of gas and dust) sputtering \citep{Draine1979b}. For simplicity, there are no grains {pre-existing} in the ICM. 
{The effects of intermixing between} dust populations that originated from cloud destruction and pre-exist in the medium will be studied elsewhere. 

We choose a computational domain as large as the material (both gas and dust) of a crushed cloud that 
 remains inside it till the end of the calculation, i.e., $t=10t_{cc}$. The cell size is 3.2~pc as a fiducial value, which corresponds to its relation to the initial cloud radius of $d_{cell}/r_{c,0} = 1/30$ that is more than three times greater than the minimum ratio $d_{cell}/r_{c,0} = 1/8$ needed for converged mass growth and cloud entrainment \citep{Gronke2018}. We also run simulations for the ratios of 1/10 and 1/20. 

We use our Eulerian gasdynamic code \citep{vns2015,vsn2017} based on the unsplit total variation diminishing (TVD) approach that provides high-resolution capturing of shocks and prevents unphysical oscillations, and the Monotonic Upstream-Centered Scheme for Conservation Laws (MUSCL)-Hancock scheme with the Harten--Lax--van Leer-Contact (HLLC) method {(see,} e.g., \citep{Toro2009}) as an approximate Riemann solver. This code has successfully passed the whole set of tests proposed in \cite{Klingenberg2007}. Simulations are run consistently with tabulated nonequilibrium cooling rates fitting the calculated ones for a gas that cools isochorically from $10^8$ down to 10~K \citep{v11,v13}.

For the dynamics of dust particles, we use our implementation of the method, in which dust grains are evolved as macroparticles {(see} description and tests in {Appendix}~\ref{AppA}~\citep{vs2024}). This method is similar to that proposed by \citet{Youdin2007}, \citet{Mignone2019}, and \citet{Moseley2023}. 
The backward reaction of dust on to gas due to momentum transfer, work conducted by the drag force and the frictional heating from dust particles, is also accounted for in order to ensure both dynamical and thermal self-consistency. 

\textls[-15]{Lagrangian particles are used as tracer particles that follow the fluid flows passively and provide information about the state of their surrounding fluid elements {(e.g.,}  \citep{Federrath2008,Vazza2010,Dubey2012}).} We consider the dynamics of Lagrangian particles as an ensemble of particles whose velocity is equal to the Eulerian gas velocity in the cell where a particle is {located}  \citep{Vaidya2018,vs2024}:
\be
 {d\pmb{x}_p \over dt} = \pmb{v}_g
\label{trace-evol}
\ee
where $\pmb{v}_g$ is the gas velocity vector in the cell where a particle is located. We solve this system coupled with the gasdynamic equations. 

In our simulations, we put both dust and Lagrangian particles inside a cloud only. The number of particles of each type in the run with low resolution (the ratio $d_{cell}/r_{c,0} = 1/10$) is 8 million, while in the run with high/middle resolution ($d_{cell}/r_{c,0} = 1/30$ and 1/20) it is increased to 32 million.

\subsection{Tracking Methods}

In order to study how the cloud material mixes with the inflow we need to separate it from the ambient gas. To do this, we use two methods. The first is based on the evolution of a Lagrangian tracer concentration or color fluid \citep{Xu1995} (below we refer to it as `colored' fluid). We assume that the value of the tracer concentration is $C=1$ inside the cloud and zero elsewhere outside it. The concentration variable is updated with
the same numerical scheme as the other fluid variables. This approach is commonly used to trace colored fluid, although it does not avoid numerical diffusion {(see} discussion in, e.g., \citep{Xu1995,Avillez2002}). The mass of the colored fluid that is associated with the cloud material depends on the choice of a level of the concentration used to constrain the cloud. Usually, this value is taken as being equal to $C=0.1$.

In another approach, Lagrangian trace particles are useful when considering entrainment processes at turbulent/laminar interfaces {(see} for review \citep{Toschi2009}). Velocity and density distributions shown by trace particles are highly intermittent {(e.g.,} \citep{Konstandin2012}). There are many applications in which the transport or aggregation of particles in turbulent flows is important: the mixing and transport properties of the ICM \citep{Vazza2010}, mergers of galaxy clusters \citep{Dubey2012}, dense parts of molecular clouds \citep{Federrath2008}, and so on. In general, trace particles are useful in any astrophysical problem in which the processes of mixing or advection are investigated; for these purposes, the dynamics of Lagrangian particles has been implemented in several public codes: ENZO \citep{Vazza2010}, FLASH \citep{Dubey2012}, AREPO \citep{Genel2013}, Pluto \citep{Vaidya2018}, etc.

\section{Cloud Tracking}
\label{sec-res0}

\begin{figure}[H]
\includegraphics[width=6.5 cm]{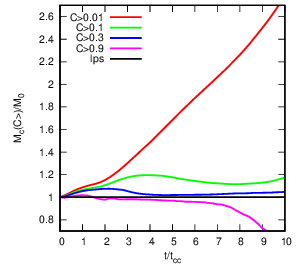}
\includegraphics[width=6.5 cm]{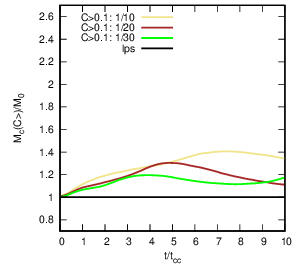}

\caption{
{\it {Left}}. {Mass} fraction of colored gas during the crushing of a cloud by a shockwave for the tracer concentration levels $C > 0.01$, 0.1, 0.3, 0.9 (color lines).
{\it {Right}}. Mass fraction of colored gas for the runs with different spatial resolution (the ratio of the cell size to the cloud radius {is shown in the legend}) for $C > 0.1$.
The mass of gas tracked by Lagrangian particles is shown by black line: it remains constant.
The masses are normalized to the initial mass of a cloud $M_0$.
The time is in units of crushing timescale $t_{cc}=3$ Myr for the parameters adopted here.
}
\label{fig-mcolor}
\end{figure}

Figure~\ref{fig-mcolor} shows the mass fraction of the colored gas for different levels of the tracer concentration $C$: $M_c(C>)/M_0$ (left panel). As defined, the concentration is $C=1$ inside the cloud at the initial moment. {During evolution, the tracer enclosed initially inside the cloud mixes with ambient noncolored gas due to numerical diffusion. This makes such gas belong to the cloud. Till $t\sim t_{cc}$, the mass of the colored gas with level $C>0.9$ remains close to the initial mass of the cloud. Whereas} for $C>0.1$ the mass of the colored gas is increased by less than 5\% to the initial mass of the cloud, for a lower value of $C>0.01$, it is about 10\% {at this time}. During further evolution, the ratio $M_c(C>)/M_0$ for $C>0.9$ stays around the same 
 until $t\sim 3t_{cc}$ and begins to drop gradually until $\sim$$7t_{cc}$, and then even more substantially. The ratio for $C>0.1$ {varies} weakly and increases not more than 20\% till the final time {$10t_{cc}$}, {while} the ratio for the lowest level $C>0.01$ increases monotonously at a factor of more than 2.6 {at this time}. 

The increase in numerical resolution leads to better tracking of cloud material. {The right panel in} Figure~\ref{fig-mcolor} presents the mass fraction of the colored gas for the runs with different spatial resolutions (the ratio of the cell size to the cloud radius) for $C > 0.1$. As expected, the ratio becomes closer to being the same as that {for a higher resolution}, which requires larger {computational} resources. 

Using trace particles, we follow the gas in a cloud. Lagrangian particles (lps) can trace the gas because their velocities are kept equal to that of the gas in a cell, where a particle is located, at each instance. It is obvious that the number of particles enclosed in a cell should be large enough for reasonable tracking of the gas in the cloud. Usually, this number is about ten particles per cell {(e.g.,} \citep{Vazza2010}). In our task, a cloud can be destructed and its material can be spread over a larger volume. Thus, we use more particles in our runs.

Each trace particle is ordered to trace a mass fraction of a cloud equal to $M_0/N_L$ (where $N_L$ is the total number of the particles). Thus, the total mass of these particles remains constant by definition, which can be seen in Figure~\ref{fig-mcolor}. 

Further, we consider how such particles trace the thermal phase of a gas belonging initially to a cloud. We assume that a particle carries the information of a part of the cloud material in which it is located. 

One should note that the increase in the colored gas mass can be considered as mixing of the `cloud' material with the ambient gas. In this sense, such an increase is the result not only of partial mixing with the intercloud medium but also numerical diffusion. A cloud crushed by shock or wind passes through a mixing phase {(e.g.,} \citep{Nakamura2006}); its duration is determined by the Kelvin--Helmholtz instability. However, cooling can start to play a role in the dynamics of a gas behind the shock front. Its influence depends on the metallicity, shock velocity, and external heating sources. Under such conditions, the cooling can become efficient and the development of the Kelvin--Helmholtz instability is suppressed. Thus, it is interesting to exclude these mixing effects as much as possible. Lagrangian particles provide sharper gradients and more intermittent distribution of density fields {(e.g.,} \citep{Toschi2009,Konstandin2012}).
Based on this, we apply trace particles for tracking cloud material. Another aim of using such particles is to compare their dynamics with dust particles.

\section{Dynamics}
\label{sec-res}

Let {us} follow the evolution of the diffuse cloud that interacted with {a} hot postshock inflow. As mentioned above, after {the} interaction {begins}, the shock front moves into the cloud with {the} velocity {of factor} $\chi^{1/2}$ lower. {This} corresponds to 30~km~s$^{-1}$ for the model considered here. 
{The} gas behind the front in the cloud is heated to $T\sim 10^5$ K. {The} cooling time for the metallicity 0.1$\zsun$ is about 30~kyr, which is significantly shorter than the crushing time $t_{cc}\sim 3$ Myr.
{Such an efficient cooling prevents the stripping of outer layers of the cloud. Indeed, the stripping operates under the action of the Kelvin--Helmholtz instability with the characteristic time $t_{KH}\sim \chi^{1/2} \lambda/v_{c,w}$, where $\lambda$ with the maximum growth rate is comparable to the cloud radius $\lambda\sim r_{cl,0}$ as noted by \citet{Klein1994}. It is seen that the time} needed for developing the longest scale of the Kelvin--Helmholtz instability coincides with the crushing time. {Therefore, one can assume that the cloud compressed by the shock front and the postshock flow does not experience a considerable mass loss due to stripping at timescales shorter than the crushing time $t\simlt t_{cc}$. }
One can see that the Kelvin--Helmholtz instability is suppressed within several crushing times under the conditions considered here. However, it can become significant again; certainly in the case where the compressed lump does not come to be gravitationally unstable. 
A short discussion about the importance of cooling can be found in {Appendix}~\ref{AppA}, where the interaction of an adiabatic cloud with a shockwave is described.



\begin{figure}[H]
\includegraphics[width=3.8cm,angle=270]{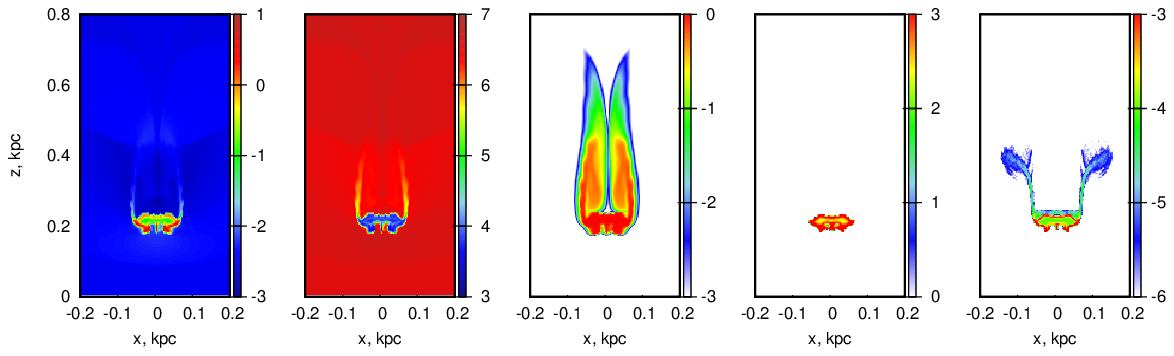}
\includegraphics[width=3.8cm,angle=270]{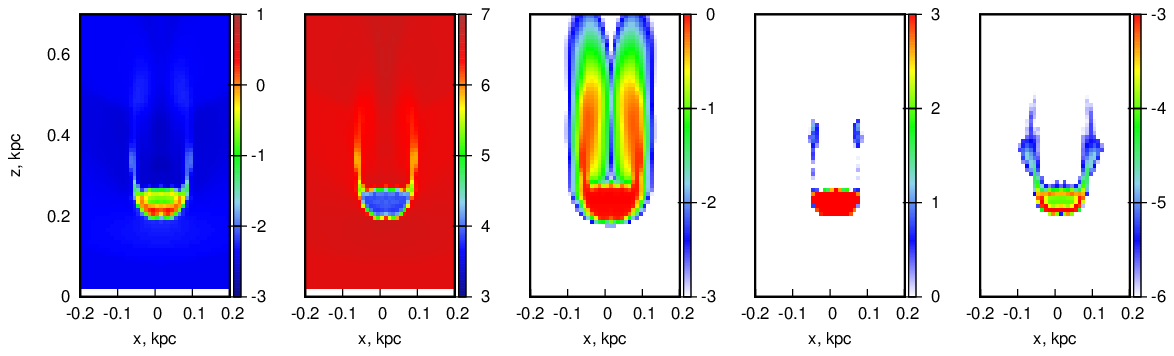}

\caption{
Density and temperature of gas, color, density of Lagrangian particles, density of dust particles in logarithmic scale ({from left to right panels}) at $t=2.4$ Myr $=0.8t_{cc}$. The upper panels present the slices for the fiducial model (the highest resolution) with  $d_{cell}/r_{c,0} = 1/30$, and the lower panels show the maps for the model with the lowest resolution $d_{cell}/r_{c,0} = 1/10$.
}
\label{fig-maps}
\end{figure}

Figure~\ref{fig-maps} presents the distributions of various values for gas and dust at \linebreak  $t=2.4$ Myr $=0.8t_{cc}$. Here, we describe in detail the slices for the fiducial model with $d_{cell}/r_{c,0} = 1/30$ (the upper row of panels). Before we proceed to this description, we note that the morphology of the cloud material depends slightly on the spatial resolution. In the lower panels, one can find the maps for the model with the lowest resolution $d_{cell}/r_{c,0} = 1/10$. It is worth noting that the morphology of the cloud as a whole stays the same. The increase in resolution leads to more detailed structures. In the previous studies, the minimum ratio of $d_{cell}/r_{c,0} = 1/8$ needed for converged mass growth and cloud entrainment was established \citep{Gronke2018}. We use three times higher resolution for our fiducial model. In addition, the destruction of an adiabatic cloud is presented in {Appendix}~\ref{AppA}.  We carried out several runs for different spatial resolutions and numbers of trace/dust particles and found a good convergence.

One can assume that the extended structures on the edges of the compressed cloud seen in the gas density field for $t=2.4$ Myr $=0.8t_{cc}$ (first panel of Figure~\ref{fig-maps}) are connected with the stripping. Similar structures can be found in the gas temperature field (second panel). However, the estimates presented above are not in favor of stripping. We argue that inflowing gas interact{ing} with the cloud flows around it and forms such structures. During this process, the {ambient} gas mixes numerically with the {cloud represented here by colored gas} (third panel). As a result, up to $t=2.4$ Myr $=0.8t_{cc}$ the {cloud gas traced by color} forms a cometary tail extended up to $\sim$$0.5$ kpc or about $\sim$$5$ times larger than the initial radius of the cloud. 
Note that to follow the evolution of such structures till the final time $t=10t_{cc}$, we extend the grid size up to $L= 10 v_s t_{cc} = 10$~kpc, which corresponds to the number of cells equaling 3000 in the direction of the flow. We present the distributions for the earliest time, and mainly, the cloud material extends highly along the flow propagation. 
 Moreover, this time is of importance for understanding the relation between the cooling and Kelvin--Helmholtz timescales.

{At the same time,} the cloud material tracked by Lagrangian particles does not demonstrate any cometary tail (fourth panel): the {cloud `represented' by Lagrangian particles} is compressed to a pancake-like structure. Their velocity field follows the gas by definition. Initially, they are attributed to the gas belonging to the cloud, and there are no trace particles outside it. Thus, under the suppression of the Kelvin--Helmholtz instability due to efficient cooling, the gas associated with the cloud is compressed into a pancake-like structure.

Dust particles show different dynamics; they represent a separate inertial 'fluid' interacting with gas by drag force, which is responsible for the gaining velocity of dust particles.
The dynamics of the dust particles is determined by the stopping time $\tau_s \sim \rho_m a / \rho_g v_r$ \citep{Epstein1924} ($\rho_m$ is the density of the grain material, $a$ is the grain size, $\rho_g$ is the gas density, $v_r$ is the relative velocity  between gas and dust). The time needed to accelerate dust particles to the velocity of the postshock inflow is around $100$ kyr. Within this timescale, dust swept out from the outer layers of the cloud gains {the} velocity close to the postshock inflow (fifth panel of Figure~\ref{fig-maps}). {This time appears to be} longer than the cooling time, {during which the gas inside the cloud becomes colder and denser, which leads to an increase in the stopping rate of dust $\nu_s \sim \tau_s^{-1} \sim \rho_g$, and as a consequence}, the {amount of} {swept-out dust} is {apparently} low.

\begin{figure}[H]
\includegraphics[width=6.5 cm]{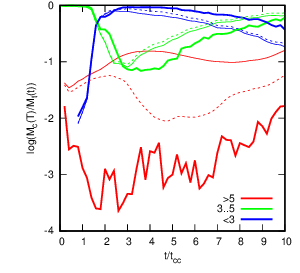}
\includegraphics[width=6.5 cm]{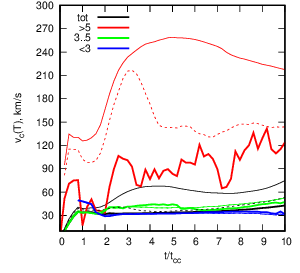}

\caption{
{\it {Left}}. Mass fractions of a gas in the cloud tracked by colored fluid for $C > 0.1$ (thin {solid} lines), {$C > 0.3$ (thin dashed lines)}, and Lagrangian particles (thick lines) for different thermal phases (the labels correspond to the logarithm of gas temperature). The mass of gas is normalized to the total mass of the cloud at the current time: $M_t(t)$. 
{\it {Right}}. {The averaged (mass-weighted) velocity for the mass fractions of a gas depicted in left panel.}
}
\label{fig-lps}
\end{figure}

Figure~\ref{fig-lps} (left panel) presents the mass of a gas {of} different thermal phases (the labels correspond to the logarithm of gas temperature). These phases are tracked by colored fluid for $C > 0.1$ (thin solid lines), $C > 0.3$ (thin dashed lines), and Lagrangian particles (thick lines). {The gas belonging to} the cloud has a temperature within $T\sim 10^4$--$10^5$~K during the first period of $\sim$$(1$--$1.5)t_{cc}$. {After that}, the major part of the gaseous mass cools down and becomes enclosed in the cold phase with $T\simlt 10^3$ K. This process is completed {at} $t\sim 3t_{cc}$. It is worth noting that two methods for tracing cloud material give very close values for the mass fractions enclosed in the cold ($T\simlt 10^3$ K) and diffuse ($T\sim 10^3$--$10^5$ K) phases. In addition, {the mass fractions} of the cloud material located in these phases by colored fluid for two levels $C > 0.1$ and $C > 0.3$ almost coincide with each other. A similar behavior can be seen for the averaged (mass-weighted) velocity {of both cold and diffuse gas (right panel of Figure~\ref{fig-lps})}. During the first $\sim$$0.5 t_{cc}$, the velocities of these phases increase almost linearly up to~$\sim$$30$ km/s and then remain around this value. 

A significant discrepancy between values determined by using two methods for tracking cloud material is found for the hot ($T>10^5$ K) phase: it reaches more than one order of magnitude for both mass fraction (compare red lines at left panel) and velocity (see right panel). According to the estimates of the cooling and Kelvin--Helmholtz timescales made above, the stripping is not efficient in the model considered here. Within $10t_{cc}$, the mass fraction of the hot gas remains around 0.1 for $C>0.1$ and drops several times for $C>0.3$. Apparently, the colored gas in the hot phase originates from numerical diffusion. This strong dependence on the level of color is in favor of numerical mixing between the gas of the cloud represented by colored fluid and the gas flowing around the cloud.

Lagrangian particles keep the total mass of the cloud exactly by definition (Figure~\ref{fig-mcolor}). One should note that, when we interpolate fluid velocity at the Lagrangian particle location, we use a shape function (here the “Cloud-In-Cell” function is chosen), so that each particle can give a nonzero contribution not only to the computational zone hosting the particle, but also to its left and right neighbors {(see, e.g.,} \citep{Youdin2007,Mignone2019,vs2024}). {This} can be considered as numerical mixing as well. {However, the} contributions are added only to the neighboring cells and {are} not spread to other ones. Under the conditions considered here, the temperature of the gas belonging to the cloud is below $10^4$ K due to efficient cooling, {and hence}, a low amount of gas can be found in the hot phase due to the interpolating procedure: this value barely exceeds 0.01. 
 Note that this mass fraction of the hot phase tracked by Lagrangian particles is much lower than that traced by colored fluid. The velocity of such {hot} gas is significantly lower than the unperturbed flow because in the interpolation {only}, the contributions from the neighboring cells are included. This picture takes place until the destruction of the cloud is inefficient.



At $t\sim (3$--$5)t_{cc}$, the mass fraction of the cold phase reaches its maximum. Afterwards, it begins to decrease apparently due to the destruction of the cold phase, which leads to the growth of the fraction in the diffuse phase. 
This comes from that the Kelvin--Helmholtz instability, which begins to play a role again. 
 At this time, the largest size of the dense {lump}\endnote{One should mention that the shape of the cloud becomes like a pancake, see Figure~\ref{fig-maps}.} decreases by $\sim$$3$ times, and the overdensity $\chi$ reaches $\sim$$100$. 
 Thus, the instability can develop at $t\simgt 3t_{cc}$.
One can note a remarkable discrepancy in both the cold and diffuse fractions after $t\simgt (2.5$--$3)t_{cc}$ between the two methods of tracking clouds: it reaches a factor of $1.5$--$2$. {It should be stressed that the fraction of the cold gas in the case of tracking by colored fluid decreases faster compared to that tracked by Lagrangian particles. }

For the cloud tracked by colored fluid, the cold fraction begins to drop just after $t\sim 3t_{cc}$. This is accompanied by the growth of {the diffuse fraction of the cloud from $\sim$$30$\% at $\sim$$5t_{cc}$ to more than 70\% at $10t_{cc}$}. {As the considerable fraction of the cloud mass transforms into the diffuse phase, it can give a wrong impression that the cloud is destroyed.} 

{If the gas of the cloud is tracked by particles, one can see that the cloud keeps its identity longer. Only after $t\simgt 5t_{cc}$ does the cold fraction {decrease} with {simultaneous} growth of both the diffuse and hot fractions. The cold fraction is higher than that determined by using the tracking by colored fluid.}

At $t \sim 10t_{cc}$, almost 99\% (by mass) of the gas belonging initially to the cloud moves with a {velocity of $\sim$$30$ km~s$^{-1}$; that is around 10\% of the unperturbed} postshock wind velocity. {This part of the cloud} is located in the cold and diffuse phases (Figure~\ref{fig-lps}). This {result} reflects {a} low entrainment efficiency {when approaching} this time. This conclusion is almost independent of the method of tracking cloud. 

\begin{figure}[H]
\includegraphics[width=6.9 cm]{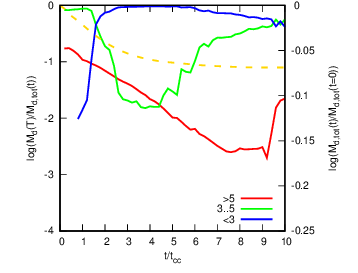}
\includegraphics[width=5.8 cm]{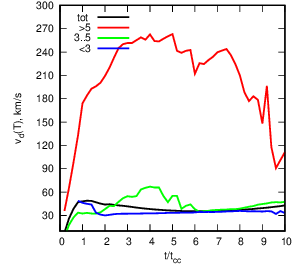}

\caption{
{\it {Left}}. Mass fractions of dust contained within a gas of the different thermal phases, $M_{d}(T)/M_{d,tot}(t)$ ($M_{d,tot}(t)$ the total mass of dust in a cloud at the current time). {The notations and line styles are the same as in Figure \ref{fig-lps}.} The dust mass fraction $M_{d,tot}(t)/M_{d,tot}(t=0)$ depicted by dashed line is shown in right axis. 
{\it {Right}}. The averaged (mass-weighted) velocity for the mass fractions of dust depicted in left panel.
}
\label{fig-dps}
\end{figure}

The mass fractions of dust particles contained within a gas in both the diffuse and cold phases evolve closely to these Lagrangian particles (compare left panels of Figures~\ref{fig-lps} and \ref{fig-dps}). 
At the same time, in the hot phase, the dust particles demonstrate different dynamics (Figure~\ref{fig-dps}). {This is because} they are swept off by the shockwave from the sides of the cloud into {the} hot gas, where they become involved in the hot postshock flow (fifth panel of Figure~\ref{fig-maps}). One can see that the velocity of the dust associated with {the} hot phase rapidly increases up to $\sim$$200$ km/s during $\sim$$(1$--$2)t_{cc}$, and after $t\sim 3t_{cc}$, reaches $\sim$$250$ km/s, which is $\sim$$0.8$ of the wind velocity. Under such conditions, dust particles are efficiently {destroyed} by sputtering. 
{This} determines the mass loss of dust: the decrease in the total {dust} mass fraction is about 10\% up to $t\sim 3t_{cc}$ ({as} shown by the right axis of the left panel in Figure~\ref{fig-dps}). 
During the later evolution, the mass-averaged dust velocity {decreases} due to destruction of the cloud and, as a consequence, replenishes the hot phase with still {undestroyed} dust particles {stripped} from the cloud and with lower velocities. 



\begin{figure}[H]
\center{\hspace{-32pt}
 \includegraphics[width=4.4 cm]{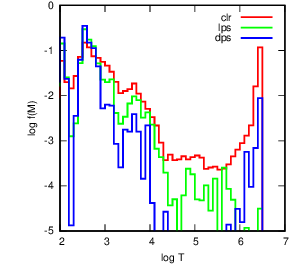}
\includegraphics[width=4.4 cm]{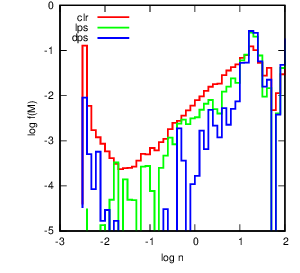}
\includegraphics[width=4.4 cm]{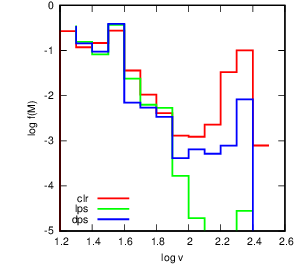}\\
\hspace{-28pt}
\includegraphics[width=4.4 cm]{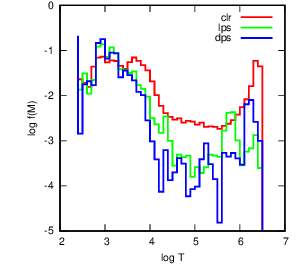}
\includegraphics[width=4.4 cm]{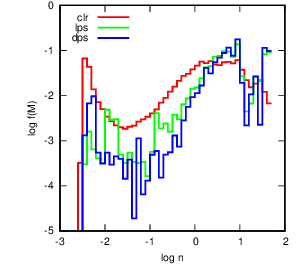}
\includegraphics[width=4.4 cm]{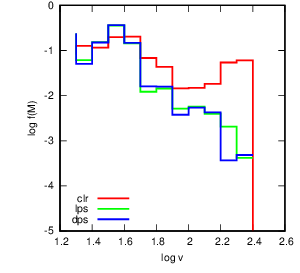}
}
\caption{
Mass fraction distribution functions of temperature (left), number density (middle), velocity (right) of a cloud material tracked by colored fluid (red lines, {labeled 'clr'}), Lagrangian particles (green lines, {labeled 'lps'}), and dust particles (blue lines, {labeled 'dps'}) at time moments {$5t_{cc}$ (top row), $10t_{cc}$ (bottom row)}.
}
\label{fig-pdf100}
\end{figure}

Above we consider the values averaged over {a few coarse} thermal bins. {In order to} take a look at more detailed distributions, we choose two time moments: 5 and $10t_{cc}$. Figure~\ref{fig-pdf100} present the mass fraction distribution functions of  temperature, number density, and velocity of a gas in a cloud for these moments. {The} two peaks clearly seen in all distributions correspond to the cold dense clump and to the hot rarefied flow, respectively. The fraction of a gas {in the intermediate range} is smaller than those enclosed in both the cold and hot phases, but it increases with time. The distributions {of the cloud fractions} tracked by colored fluid demonstrate a {considerable} excess over those tracked by Lagrangian particles; this becomes more remarkable for the diffuse gas with $T\simgt 5\times 10^3$ K and $n\simlt 1$ cm$^{-3}$ moving with $v\simgt 50$ km/s. This discrepancy increases at $10t_{cc}$ in comparison with $5t_{cc}$ and will continue to grow during further evolution due to ongoing cloud destruction.

At both time moments, the mass distributions for Lagrangian and dust particles demonstrate {similar behaviors for} the cold phase only. {This means that} Lagrangian particles {provide an efficient tracing of dust} associated with only cloud material in the cold phase. The mass fraction of dust associated with hot gas is {less than several percent}. Such dust particles are of importance due to their efficient sputtering: at first, they can enrich the ambient gas with metals and, secondly, by becoming smaller {in size} they emit in a short infrared range. {In the range $T\sim 10^4$--$10^5$ K, there is a significant gap in the distribution for the gas in which dust particles are enclosed at $5t_{cc}$ (blue line, upper left panel of Figure~\ref{fig-pdf100}). More exactly, the mass fraction in this range is less than 0.01\%. Due to sweeping, this fraction increases more than 100 times up to the time $10t_{cc}$ (lower left panel). The same can be found for the intermediate values of the number density of the gas where the dust particles are located: one can see that the mass fraction of the dust associated with the gas of $n\sim 0.01$--$0.1$ cm$^{-3}$ is also less than 0.01\%, and at $t=10t_{cc}$, the fraction increases by 30--100 times (see the two middle~panels).}

At $t = 5t_{cc}$, one can see the considerable differences in the velocity distributions for $v\simgt 50$ km/s, which become more significant for $v\simgt 100$ km/s (upper right panel). During further evolution, the discrepancy between the Lagrangian and dust particles decreases, whereas the distribution of the gas velocity tracked by colored fluid keeps the significant difference from other distributions.



\section{Conclusions}
\label{sec-con}

In this work, we analyzed how a diffuse radiatively cooled cloud ($\chi=100$) interacts with the hot $300$ km/c postshock wind. We utilized two methods to track the cloud material: the first is based on a ‘colored’ fluid approach, and the second one uses the Lagrangian (trace) particles. We followed the dynamics of the dust grains interacting with the ambient gas by the drag force up to their destruction after they are swept out of the ‘parent’ cloud by the hot flow. We find that the radiative losses can play a significant role in the dynamics of a gas behind the shock front.  We find that under the conditions considered in this work, the cooling becomes efficient during the earliest times and the development of the Kelvin--Helmholtz instability is suppressed. In particular, the destruction of a warm cloud of size 100 pc with metallicity higher than 0.1 $\zsun$ interacting with the hot flow of the ICM is expected to be delayed up to several crushing times for such a cloud.

For a gas tracked by colored fluid, we show that the total mass of a gas associated with the cloud strongly depends on the level of tracer concentration higher than that of the gas that is believed to belong to the cloud: 
 for a low level, $C>0.01$, the discrepancy reaches a factor of several times at ten crushing times of the cloud ($10t_{cc}$); the increase in the level results in a smaller difference but  leads to unpredictable variations in the total mass estimated by using colored fluid during the evolution. Certainly, the increase in spatial resolution a gives lower difference. Such an increase in the colored gas mass can be represented by mixing of the `cloud' material with the ambient gas. This process includes unavoidable effects of numerical diffusion.

We find that when traced by Lagrangian particles, the cloud material has sharper gradients and more intermittent density field distribution than when traced by the colored fluid.  However, it provides an alternative approach to trace the cloud material.
There is a special interest in this method because trace particles are used for modeling dust \citep{Farber2022}.

We found that at $t\sim 10t_{cc}$ almost 99\% (by mass) of the gas, initially attributed to the cloud, has at most 10\% of the postshock wind velocity. We interpret this as being a result of the weak entrainment efficiency of the cloud. We find that, irrespective of the approach used to trace the cloud material, such gas is enclosed in the cold  ($T\simlt10^3$ K) and diffuse ($T\sim 10^3$--$10^5$ K) phases. However, the gas fractions in the cold and diffuse phases differ depending on the method: the colored fluid approach results in about two times lower mass being found in the cold phase.

Our results suggest that the mass fraction of the cloud found in the hot phase ($T > 10^6$ K) depends on the tracer concentration level. This fraction appeared to be significantly higher for the fiducial level of $C > 0.1$ than that determined in the case when the  Lagrangian particles are used to trace the gas of the cloud.

Using tracking methods, one can infer the cloud entrainment efficiency, and the distribution of cloud material between various thermal phases. The two methods considered in this work allow the detection of differences in the dynamics of the cloud material. In particular, they reveal variations in the gas distribution in each of the assumed thermal phases, as well as distinctive mixing of the cloud material with the ambient gas. We note that these methods depend on spatial resolution: the accuracy with which the cloud material is traced increases for a smaller cell size and/or larger number of trace particles.

Dust particles can be swept out from their 'parent’ cloud material; further, their dynamics in diffuse and hot rarefied gas differs from that of Lagrangian particles. In dense parts of a cloud, the velocity of the shockwave is significantly lower, so the dust dynamics as an ensemble of particles is close to that of Lagrangian ones. However, the difference becomes more prominent in the cloud parts mixed efficiently due to an irregular velocity field. In addition, the dynamics of dust particles is changed by their destruction, which becomes significant in the hot, high-velocity gas.  Therefore, the dust particles cannot be correctly described as Lagrangian ones in a wide range of conditions that appear in the process of the interaction between a cloud and a hot wind (or strong shockwave).

The destruction of dusty clouds interacting with a hot flow is closely related to the accurate interpretation of the extinction measurements in the ICM. The large scatter of the extinction values can be connected not only with observational uncertainties {(e.g.,} see \citep{Shchekinov2022Univ,Muller2008} for a review), but also with erroneous conclusions about the destruction of dusty clouds moving in the hot gas.

\vspace{6pt} 

\authorcontributions{The authors have contributed equally to this work. All authors have read and agreed to the published version of the manuscript.
}

\funding{Numerical simulations of the cloud dynamics were supported by the Russian Science Foundation (project no. 23-22-00266).
}

\informedconsent{Not applicable. 
}

\dataavailability{The data underlying this article are available in the article. 
}
\acknowledgments{We thank Yuri A. Shchekinov for their many valuable comments and Ilya Khrykin for several suggestions. We are grateful to the anonymous referees, whose comments helped us to improve this~paper.
}
\conflictsofinterest{{ The authors declare no conflict of interest.} 
} 

\abbreviations{Abbreviations}{
The following abbreviations are used in this manuscript:\\

AGN---Active Galactic nucleus; ICM---intracluster medium; ISM---interstellar medium, CGM---Circumgalactic Medium; IGM---Intergalactic Medium.
}

\appendixtitles{yes} 
\appendixstart
\appendix
\section[\appendixname~\thesection]{Tracking an Adiabatic Cloud}\label{AppA}

Let us consider the evolution of an adiabatic cloud with $n_c = 10$ cm$^{-3}$, $T_c = 10^2$ K embedded into the medium with $n_m = 0.1$ cm$^{-3}$, $T_m = 10^4$ K, i.e., the overdensity is $\chi=100$. The radius of the cloud is equal to 4pc. The inflow velocity of a gas $v_s$ is set to 300 km/s. The crushing time is $\sim$100~kyr for these parameters. The postshock flow has parameters according to the Rankine--Hugoniot conditions. The cell size is 0.15~pc, that corresponds to its relation to the initial cloud radius of $d_{cell}/r_{c,0} = 1/32$.

\begin{figure}[H]
\includegraphics[width=10cm]{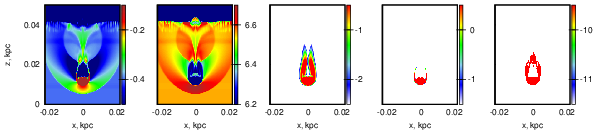}
\includegraphics[width=10cm]{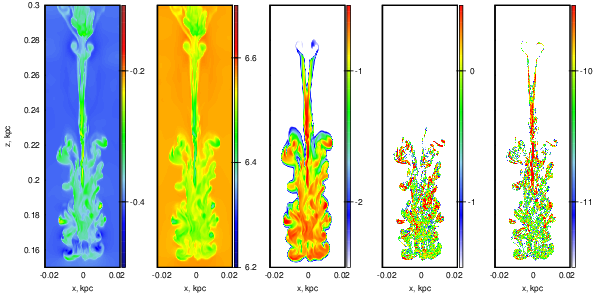}

\caption{Density and temperature of gas, color, density of Lagrangian particles, density of dust particles in logarithmic scale (from {left to right panels}) at $t=1t_{cc}$ ({top panels}) $t=10t_{cc}$ ({lower~panels}).}
\label{fig-maps-a}
\end{figure}

Figure~\ref{fig-maps-a} shows the distributions for the adiabatic cloud at times $t_{cc}$ (top) and $10t_{cc}$ (bottom). One can resolve the typical evolution of the cloud interacted with the shockwave {(compare to e.g.,} \citep{Nakamura2006}). Already at early time there is a difference for the density (first panel) and the color (third panel) distribution. The latter is more diffused in comparison with the distribution of Lagrangian particles, which demonstrates that the narrow structures stripped by the postshock flow have the morphology similar to the density field. Dust particles are swept out from these extended structures and show more complex distribution.



At later stages, the Kelvin--Helmholtz instability develops and the cloud transitions into the mixing phase. Till $10t_{cc}$ (bottom panels) the cloud is almost completely destructed: the average overdensity in the resulted structure is less than 0.2 dex. The temperature distribution shows the variations of the same order. The volume occupied by colored fluid is substantially larger compared to the initial volume of the cloud (one can compare the upper and lower panels). The distributions for color and trace particles are comparable only on large scales (except the extended tail along the line with $x=0$). On small scales the distribution of particles is more abrupt and intermittent. However, one can easily find many eddies---signatures of the Kelvin--Helmholtz instability. Note that the morphology of the dust distribution is close to that of trace particles, except the extended tail along the line with $x=0$. The dust-to-gas ratio is less than $10^{-8}$ at $10t_{cc}$ ($10^{-3}$ at the initial moment), which corresponds to complete destruction of dust. For completeness, we do not remove the dust particles reaching the smallest size, i.e., 10 \AA. Such small grains follow the gas velocity field. Thus, the similar morphology is expected at later evolutionary stages, which is apparent in Figure~\ref{fig-maps-a}.



\begin{figure}[H]
\includegraphics[width=6.5 cm]{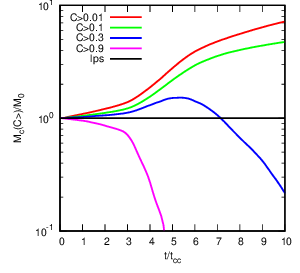}
\caption{Mass fraction of colored gas during the crushing of an adiabatic cloud by a shockwave for the tracer concentration levels $C > 0.01$, 0.1, 0.3, 0.9 (color lines).
The mass of gas tracked by Lagrangian particles is shown by black line: it remains constant.
The masses are normalized to the initial mass of a cloud $M_0$.
The time is in units of crushing timescale $t_{cc}=100$ kyr for the parameters adopted here.
}
\label{fig-mcolor-a}
\end{figure}

Figure~\ref{fig-mcolor-a} shows the mass fraction of colored gas for different levels of the tracer concentration $C$ for the crushing of an adiabatic cloud by a shockwave. The ratio for the lowest level $C>0.01$ increases monotonically by a factor of more than 7 times till $t=10t_{cc}$. Even for higher level $C>0.1$ the mass growth reaches about a factor of 5. As mentioned previously this can be considered as mixing of the "cloud" material with the ambient gas. That suggests a highly efficient development of Kelvin--Helmholtz instability in the absence of any radiative losses (see for comparison the left panel of Figure~\ref{fig-mcolor}). Note that this mixing cannot be differentiated from numerical diffusion.

Radiative losses can influence the evolution significantly. At low cooling rate the development of Kelvin--Helmholtz instability goes as usual. Density gradients are low, which does not affects on formation of eddies. Increasing the rate leads to faster cooling than the development of the eddies. At first, this proceeds at small scales. The more efficient cooling does not allow growth of the instability on larger scales. During the cloud-wind interaction the largest unstable wavelength is the initial size of a cloud, thus, there is a cooling rate (under given physical parameters of both cloud and shockwave) at which the Kelvin--Helmholtz instability should be suppressed.
\begin{adjustwidth}{-4.6cm}{0cm}
\printendnotes[custom]
\end{adjustwidth}


\begin{adjustwidth}{-\extralength}{0cm}

\reftitle{References}

\PublishersNote{}
\end{adjustwidth}

\end{document}